# Information Theory and Thermodynamics

Oded Kafri

Varicom Communications, Tel Aviv 68165 Israel.

A communication theory for a transmitter broadcasting to many receivers is presented. In this case energetic considerations cannot be neglected as in Shannon theory. It is shown that, when energy is assigned to the information bit, information theory complies with classical thermodynamic and is part of it. To provide a thermodynamic theory of communication it is necessary to define equilibrium for informatics systems that are not in thermal equilibrium and to calculate temperature, heat, and entropy with accordance to Clausius inequality. It is shown that for a binary file the temperature is proportional to the bit energy and that information is thermodynamic entropy. Equilibrium exists in random files that cannot be compressed. Thermodynamic bounds on the computing power of a physical device, and the maximum information that an antenna can broadcast are calculated.

# Introduction

Shannon information and Boltzmann entropy have the same mathematical expression. However, information is conceived as an opposite of entropy. Brillouin in his book suggested that information is negative entropy [1]. Many view information as a logical sequence of bits of some meaning as oppose to a thermal state, which is a state of randomness. The known scientific knowledge does not support this mystic idea. Shannon has shown that the higher the randomness of the bits in a file, the higher the amount of information in it [2]. The Landauer and Bennet school [3] suggests that the randomness of the bits in a file is related to Kolmogorov complexity [4]. This claim may give an impression that the Shannon information is a meaningful subjective quantity. However, according to the Shannon theory a compressed file, containing meaningful information, has similar amount of information as an identical file, with one flipped bit that cannot be decompressed and therefore, for us the receivers, it is just a noise.

In this paper a thermodynamic theory of information is proposed. It is shown that Shannon information theory is a part of thermodynamics, and that information is the Boltzmann -$H$ function. Therefore, information has a tendency to increase the same way as entropy.

The information increase observed in nature is attributed to a specific mechanism rather than to a natural tendency. Here it is proposed that increase of information as increase of thermal entropy, is caused by the second law of thermodynamics.

To support this claim it is required to calculate, for informatics systems, the quantities in Clausius inequality (which is the formulation of the second law), namely, entropy, heat, and temperature and to define equilibrium.

In section I, the classical thermodynamics of heat transfer from a hot bath to a cold bath is reviewed together with the basic definitions of entropy, heat, temperature, equilibrium, and the Clausius inequality.

In section II, a calculation of the entropy, heat, temperature, and the definition of equilibrium for the transfer of a one-dimensional two-level gas from a hot bath to a cold bath according to statistical mechanics are provided and an analogy to Clausius inequality is shown.

In section III, an analysis of the transfer of a frozen one-dimensional two-level gas (a binary file) from a hot bath (a broadcasting antenna) to a cold bath (the receiving antennas) is provided. A temperature is calculated to the antenna that, together with the transmitted file information (entropy) and its energy (heat), is shown to be in accordance with classic thermodynamics and the Clausius inequality. Therefore, It is concluded that in the absence of thermal equilibrium information is entropy

In section IV, these results are used to calculate a thermodynamic bound on the computing power of a physical device and in section V, a thermodynamic bound on the maximum amount of information that an antenna can broadcast is calculated.

**I - Classical thermodynamics of heat flow**.

Clausius deduced the second law of thermodynamics from Carnots calculation of the maximum amount of work $\Delta W$ that can be extracted from an amount of heat $\Delta Q$ transferred from a hot bath at temperature $T_H$ to a cold bath at temperature $T_C$ [5]. Carnot used in his machine an ideal gas as a working fluid, and the gas law for his calculation. The Carnot efficiency is,

$$\eta \equiv \Delta W/\Delta Q \leq 1 - T_C/T_H. \tag{1}$$

Namely, the maximum efficiency $\eta$ of a Carnot machine depends only on the temperatures. To obtain the maximum efficiency the working gas should obey the gas law, therefore, the machine has to work slowly and reach equilibrium at any time. Clausius [5] assumed that Carnot efficiency is always true no matter what mechanism or working fluid is used. That means that there is no dependence on the ideal gas law. Clausius concluded that if the Carnot efficiency is always true, there is a quantity, entropy $S$, that is defined in equilibrium (the equality sign) and can be calculated according to,

$$\Delta S \geq \Delta Q/T \tag{2}$$

When a system is not in equilibrium, $\Delta Q/T$ is smaller than the entropy. This inequality reproduces the Carnot efficiency. However, it reveals more than one would expect. The entropy change $\Delta S$ is equal to $\Delta Q/T$ only in equilibrium. Out of equilibrium $\Delta Q/T$ is smaller than entropy. Therefore if we assume that any system has a tendency to reach equilibrium, any system tends to increase $\Delta Q/T$. Clausius assumed that taking a system out of equilibrium requires work, which will also eventually reach equilibrium (namely it will be thermalized) and, therefore the entropy of a closed system tends to increase and cannot decrease. Temperature and entropy are defined at equilibrium and the temperature can be calculated as,

$$T = (\Delta S/\Delta Q)_{equilibrium} \tag{3}$$

This definition of temperature is accepted to be always true.

Now I calculate a simple example of the entropy increase in heat flow from a hot thermal bath to a cold one (see Fig 1).

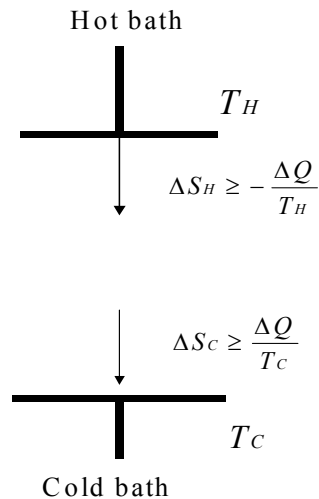

Figure 1: *The entropy increase in spontaneous energy flow from a hot thermal bath to a cold thermal bath.*

When we remove an amount of energy $\Delta Q$ from the hot bath, the entropy reduction at the hot bath is $\Delta Q/T_H$. When we dump this energy to the cold bath, the entropy increases by $\Delta Q/T_C$. The total entropy increase is $\Delta S = \Delta Q/T_C - \Delta Q/T_H$. One can see that if the process is not in equilibrium $\Delta S > \Delta Q/T_C - \Delta Q/T_H$. In general

$$\Delta S \geq \Delta Q/T_C - \Delta Q/T_H \qquad (4)$$

In sections II and III I will give an analogy to this example for statistical physics and for information theory.

**II - Statistical Physics of one-dimensional two-level gas.**

The entropy is defined in statistical physics as $k\ln\Omega$, where $\Omega$ is the probability of the system and $k$ is the Boltzmann constant [6]. I will use this definition to calculate the thermodynamic quantities and the Clausius inequality for a system that resembles an informatics system.

I consider a thermal bath at temperature $T_H$, which is in contact with a sequence of $L$ states. $p$ of the $L$ states have energy $\varepsilon$ and I name them "one". $L-p$ of the states have no energy and I name them "zero". I analyze the thermodynamics of transferring this two-level gas from a hot bath at temperature $T_H$ to a colder bath $T_C$. The probability of the two-level sequence is simple, because the number of possible combinations of $p$, "one" particles in $L$ states is the binomial coefficients (see Fig 2), namely, there are, $\Omega = L!/[\,p!\,(L-p)!]$ possible combinations.

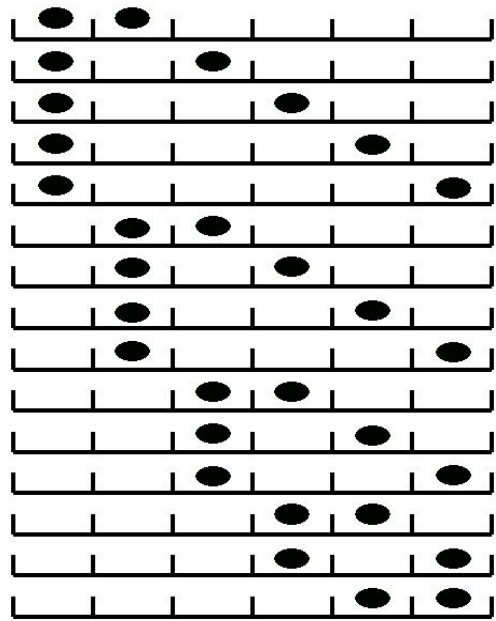

Fig 2: *One-dimension two-level gas with L=6 and p=2. In equilibrium all possible combinations have equal probability (the ergodic assumption). If some of the combinations have higher probability than others the system is not in thermal equilibrium.*

The entropy of the system is $k\ln\Omega$, and the energy of the system is $p\varepsilon$. The temperature is calculated from Eq. (3). Using Stirlings formula we derive $\partial Q/\partial S$ to obtain $1/T$. The well-known result is,

$$p/(L-p) = \exp(-\varepsilon/kT) \text{ or } T = (\varepsilon/k)/\ln[(L-p)/p]. \qquad (5)$$

Namely, one parameter $T$ represents all our knowledge on this one-dimensional two-level gas in equilibrium. This is a well-defined system with a well-defined entropy temperature and energy. The equilibrium was invoked by giving an equal probability to all the possible combinations of the $p$ particles in $L$ states (This assumption is called the ergodic assumption [7]). Eq. (5) is a famous result, but it should be noted that my derivation is done according to $Q$ (the heat) and not according to the internal energy of the gas as is done in most textbooks. The reason is that in my model, a two-level gas is transferred from a hot bath to a cold bath, and therefore its energy is heat. If a system is not in equilibrium, there are certain combinations that are preferred (a biased distribution), and thus the actual combination span (phase space) is smaller. Therefore, the probability $\Omega$ of the gas not in equilibrium is smaller. Since the energy of the gas is conserved, we obtain a higher effective "temperature". Boltzmann called the quantity $k\ln\Omega$ calculated for a biased distribution the $-H$ function [8].

When the two-level gas is removed from the hot bath, the entropy is reduced by $\Delta S_H = p_H\varepsilon/T_H = kp_H\ln[(L-p_H)/p_H]$. When we dump it to the cold bath, we generate entropy $\Delta S_C = p_H\varepsilon/T_C = kp_H\ln[(L-p_C)/p_C]$.

The total change in the entropy is,

$$\Delta S_C - \Delta S_H = (k\Delta Q/\varepsilon)\ln\{(p_H/p_C)[(L-p_C)/(L-p_H)]\} \geq \Delta Q/T_C - \Delta Q/T_H \qquad (6)$$

It should be noted that part of the energy, $\varepsilon(p_H - p_C)$, was transferred from the gas to the cold bath. Obviously Eq. (6) is positive when $T_C$ is lower than $T_H$, and we see that Eq. (6) is with accordance with Eq. (4), namely, the second law (see Fig 3). The inequality stands for the transmission of a one-dimensional two-level gas with a biased probability of the combinations of the gas. The probability in mechanical statistics is considered sometimes as a time average on all possible combinations of $p$ particles in $L$ states. However, if we look at our one-dimensional two-level gas at a given short time, we will observe only one of the possible combinations.

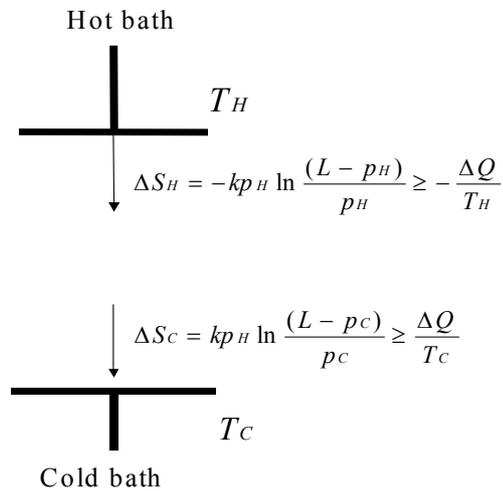

Fig 3. *The entropy increase, due to transmission of one-dimension two-level gas, from a hot bath to a cold bath.*

In fact, we will see a binary file. Adopting a slightly different point of view can solve this paradox, namely, instead of considering the system probability as a time average, we consider it as the probability of finding a given combination at a certain time. In the ergodic case, in equilibrium, any one of the possible combinations can pop up at a given time without any preference. In a biased system, not in equilibrium, certain combinations will have higher probabilities than others. This approach does not affect the mathematical analysis, however, it will be very useful when I consider information.

**III - Information theory of one-dimensional two-level sequence.**

The Shannon definition of information is based on a model of a transmitter and a receiver. In his model a binary file is transferred from a transmitter to a receiver. A binary file is, in fact, a frozen one-dimensional two-level gas. The binary file is not in thermal equilibrium as it is highly biased to one possible combination of the bits. As oppose to two-level gas, where the energy of the bit $\varepsilon$ is fixed, in a binary file the bit energy may change continuously.

Shannon was interested in the maximum amount of information that can be coded in a given binary file of length $L$. His famous result is that information has the same expression as entropy. However, no connection was made between Shannons entropy and thermodynamics. The amount $p$ of "one" bits, in a file of length $L$, has no unique relation to the amount of information in the file. This is in contradistinction to the two-level gas, in which the energy, the temperature and the entropy are functions of $p$, (see Fig. 3). For example, in several files having the same $p$ some may have a very small amount of information, i.e. if all the "one" bits are in the beginning of the file, and the rest of the file has zero bits or any other ordered combination (see Fig. 4), and some other files may have a relatively high amount of information, if the distribution of the bits in the file is random.

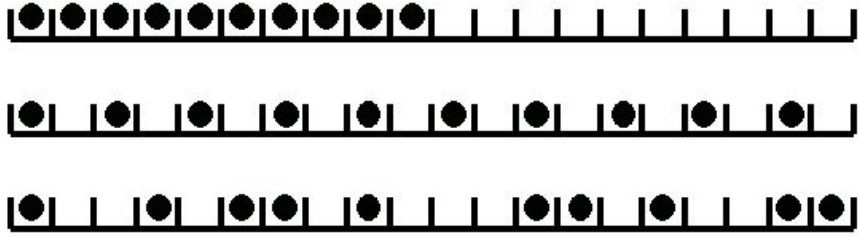

Fig 4: *Three possible binary files having the same energy. The higher two files have higher order and therefore contain little information and can be compressed effectively. The lower file is random and contains maximum information and therefore it cannot be compressed and is in equilibrium.*

The amount of information in a file is a function of the randomness of the bits in it, and there is no unique connection between the number of bits and the amount of information. The reason that Shannon obtained the same expression as Boltzmann is that in two-level gas we have no way to predict what combination will be at a certain time, and in a random file we have no way to predict what bit will be at a certain time (the unpredictable sequence of bits is the useful information). Since information and entropy are probability calculations, the same expression is obtained. Nevertheless, the calculations for the two-level gas and a file are different as I show hereafter.

With analogy to the transmission of two-level gas, I start the thermodynamic analysis of a file transmission by considering a truly random sequence of $L$ bits. In this case $p = L/2$, therefore the maximum information that a file of length $L$ can contain according to Shannon is $\Delta I = L\ln2$. In this case the ratio between the number of "one" bits (the energy) and the information (entropy) is unique. This is in contradistinction to a file with some correlations in which the number of "one" bits $p$ does not determine the amount of information. So by assigning energy $\varepsilon$ to the "one" bit we obtain $\Delta Q = L\varepsilon/2$ and $\Delta S = kL\ln2$. Using Eq.(3) we obtain,

$$T = \Delta Q/\Delta S = (\varepsilon/k)/2\ln2. \qquad (7)$$

Eq. (7) should be compared with Eq. (5) namely $T = (\varepsilon/k)/\ln[(L-p)/p]$. We can see that for a file, the temperature depends only on one variable, the bit energy. In a two-level gas the temperature depends also on $p$. In two-level gas lowering $p$ reduces the temperature and increases the entropy. So according to the second law a two-level gas would tend to cool down. In a file, reducing the bit energy yields a similar result. Therefore, according to the second law, a file has a tendency to lower its bit energy. I complete the analogy by considering antenna broadcasting a binary file to $N$ antennas. A possible deployment of such system is a point-radiating antenna surrounded by a sphere, whose area is divided to $N$ equal receivers. According to Turing model [9], the hot antenna emits the broadcasted file. A receiver antenna receives the broadcasted file but with a lower bit energy. Therefore, it is equivalent to a cold bath. Using Eq. (4) we obtain,

$$\Delta S \geq \Delta Q/T_C - \Delta Q/T_H = Nk\Delta I - k\Delta I. \qquad (8)$$

Eq. (8) shows that the file temperature obtained in Eq. (7) yields correctly the increase in information in the broadcasting of a binary file to $N$ receivers (which is $N\Delta I - \Delta I$). In "peer-to-peer" transmission, as in Shannon model, no information increase was involved; therefore no thermodynamic considerations are necessary. Out of equilibrium, there is a correlation between the bits, and the amount of information in the file is smaller. As a result, the same energy carries less information, therefore $T$ is higher and $\Delta I$ is smaller. Using Eq. (8) we can rewrite Clausius inequality for informatics system as, $\Delta S \geq k\Delta I$.

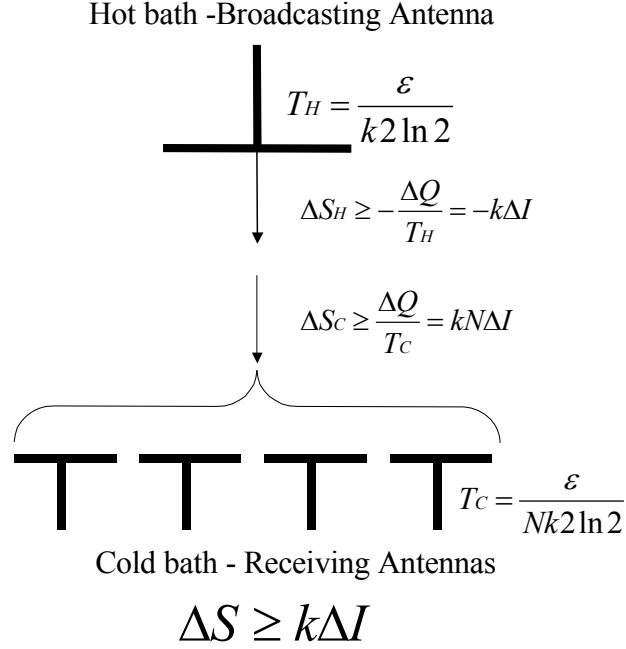

$$\Delta S \geq k\Delta I$$

Fig. 5 *The analogy between heat flow from a hot bath at temperature $T_H$ to a cold bath at temperature $T_C$ and an antenna broadcasting a file of bit energy $\varepsilon$ to N antennas, each receiving the file with bit energy $\varepsilon/N$. In the thermal case the entropy increase is $\Delta S \geq \Delta Q/T_C - \Delta Q/T_H$. However, the same equation $\Delta S \geq \Delta Q/T_C - \Delta Q/T_H$ reproduces well the information balance when we use the temperature definition from Clausius inequality $\Delta Q/\Delta S$ for a compressed binary file. The antennas deployment is drawn to emphasize the physics only.*

This implies that Information, like entropy, tends to increase. In a general case in complex systems both informatics and thermal processes occurs simultaneously. In these cases a transformation of thermal entropy to informatics entropy and vice versa may occur. Thus Clausius inequality can be written as,

$$\Delta S \geq \Delta Q/T + k\Delta I \qquad (9)$$

It is worth noting that ideally compressed files rarely exist. In order to calculate the amount of information in a file, we have to find an ideal compressor [10]. Unfortunately such a compressor does exist only asymptotically. The amount of information in an uncompressed file, with some correlation between bits, is equivalent to the Boltzmann –H function, namely the "entropy" of a system out of equilibrium with a biased distribution. Shannon, in his famous paper [2], mentioned that information is the Boltzmann $-H$ function, nevertheless it is called by many entropy.

**IV - Example -The Computing power of a Physical device**

The units used in communication are the power $P$ and the frequency $f$ in bits/sec of an emitter/transmitter and not the bit energy. Therefore, the temperature of emitter/transmitter can be written as;

$$T = P/(k f \ln 2 ). \qquad (10)$$

It is also assumed that any informatics system is surrounded by a thermal bath that emits thermal noise at a temperature $T_n$. To calculate a bound on a computing power of a physical device Turing's model [9] is used. In Turing model erasing one bit and registering it again is an example of a logical operation. In our case the bits rate of the file is the maximum computing power. The higher the bit rate, the lower the temperature of the file as the bit energy is reduced. Since the temperature of the file must be kept above the temperature of the noise $T_n$, the frequency has an upper limit. From Eq. (10) we conclude that $f \leq P/(k \ln 2 T)$ where $T$ should be about 10 times higher than the noise temperature. Therefore, the upper bound on computing power of any device is,

$$f \leq P/(10 k \ln 2\ T_n). \qquad (11)$$

The powers applied on any computing device, and its ambient temperature suffices to give a limit on its computing power. C.H. Bennett, in his review on the "Thermodynamics of computation" [3], quotes from a Von Neumann talk that "a computer operating at temperature $T$ must dissipate at least $k \ln 2 T$ per elementary act of information". Later Bennett quotes that "in nature per nucleotide or amino acid the ratio is 20-100 $k \ln 2 T$ ".

**V Example - the maximum information that an antenna can broadcast.**

Consider a point antenna broadcasting a compressed file. The bit energy at a receiving antenna having area $A$ will be lower as the distance $R$ between the transmitting antenna and the receiving antenna is higher. The temperature of the file at the receiver will be

$$T_r = T_t(A/4\pi R^2) = PA/[k \ln 2 f\, 4\pi R^2]. \tag{12}$$

The minimal size of both the transmitting and the receiving antenna is $\lambda = c/f$, where $\lambda$ is the wavelength of the carrier radiation and $c$ is the speed of light. If we assume that the bit energy at the receiver should be 10 times larger than $kT$ (a conventional assumption for signal to noise requirements), we can calculate from Eq. (12) the maximum distance $R$ that an antenna of power $P$ and a frequency $f$ can broadcast to a receiving antenna of area $A = \lambda^2$. As an example, consider a 50 W, 900 MHz radio transmitter. From Eq. (5) we find that the temperature of a broadcasted file via an antenna at the size of a wavelength is about $5 \cdot 10^{15}$ K. We can cool the file from signal to noise considerations to about 3000 K (ten times the ambient temperature). We assume that the receiver has an antenna of area $A = \lambda^2$, and we obtain that the thermodynamic bound for the maximum distance $R \approx 100$ Km. This number may appear high, however, the receiving antenna is usually linear, and $A$ is less than $1/100^{th}$ of $\lambda^2$ and thus $R \approx 10$ Km.

Now we consider a large antenna of radius $R_t \gg \lambda$. Because of the second law, it is not possible to detect a signal with a higher intensity (temperature) than that of the surface of the antenna. So we can replace $T_r$ with $T_t$ in Eq. (6), and calculate the entropy leaving the transmitting antenna. We can imagine the surface of the broadcasting antenna as a superposition of many small antennas of area $\lambda^2$ and substitute $A = \lambda^2$ in Eq. (6). The limit temperature of the broadcasted file can also be calculated from Eq. (6). The maximum information transmission of an antenna over a time interval $\Delta t$ is given by $\Delta S/k$ and yields,

$$\Delta S \equiv P\Delta t/ T_t = k \ln2 \, (f/\lambda^2) \, 4\pi R_t^2 \, \Delta t \geq k \, \Delta I, \qquad (7)$$

This is the Clausius inequality for a broadcasting antenna. This expression resembles the results of Bekenstein [11] and Srednicki [12] that state that entropy is proportional to the area of spherical emitters such as a black hole or any imaginary sphere.

**Summary**

This paper deals with the energetic of a file broadcasted from one antenna to several antennas (a generalization of Shannon's theory). An analogy between information broadcasting from one antenna to several antennas to heat flow from a hot bath to a cold bath is drawn. What is shown is that:

1. The Shannon information content $I$ of a file is equivalent to the Boltzmann $-H$ function.

2. The transmitted file energy is equivalent to heat.

3. A compressed file is a state of equilibrium.

4. The temperature of the antenna is proportional to the bit energy broadcasted from it or received by it.

Clausius inequality for an antenna and for informatics systems in general is calculated. Also, a bound on the computing power of a physical device was derived. The maximum information that can be broadcasted from an antenna was calculated, and is shown to be a function of its area.

**Acknowledgements**: I thank Y.B.Band for careful reading of the manuscript and many suggestions to R.D. Levine, J. Agassi and Y. Kafri for many useful discussions.

**References**

1. L. Brillouin *Science and Information Theory*, Academic press NY (1962) pp161.

2. C. E. Shannon "*A Mathematical Theory of Communication*", (University of Illinois Press, Evanston, Ill., 1949).

3. C.H. Bennet, Int. J Theor. Phys. 21, 905 (1982); C.H. Bennet arXiv:physics/0210005 v2 9 Jan (2003)

4. M. Li and P.M.B. Vitanyi, An Introduction to Kolmogorov Complexity and its Applications, Springer-Verlag, New York, Second Edition, 1997 (xx+637 pp).

5. J. Kestin, ed.," *The second law of thermodynamics*", (Dowden, Hutchinson and RossStroudsburg 1976) pp 312.

6. L. D. Landau and E. M. Lifshits, Statistical Physics, Pt. 1 3$^{rd}$ Edition Pg. 25 (Pergamon, New York).

7. M. Plischke and B. Bergersen, Equilibrium Statistical Physics, 2$^{nd}$ Edition Pg. 3 (World Scientific, London).

8. K. Huang, Statistical Mechanics, Pg. 68 (John Wiely, New York).

9. A. M. Turing, "On Computable Numbers with an Application to the Entscheidungsproblem", Proc. London Math. Soc. [2], **42**, 230-265, (1936-7)

10. D.A. Huffman, Proc. of the Institute of Radio Engineers **40**, 1098 (1952); J. Ziv and A. Lempel, IEEE Transactions on Information Theory **23**, 337 (1977).

11. J. D. Bekenstein, Phys. Rev. D**7**, 2333(1973).

12. M. Srednicki, Phys. Rev. Lett. **71**, 666 (1993).